\documentclass[12pt]{iopart}

\usepackage{graphicx}
\graphicspath{ {./images/} }
\usepackage{hyperref}

\begin{document}

\title[Thermal annealing of sputtered Nb$_3$Sn and V$_3$Si thin films for superconducting RF cavities]{Thermal annealing of sputtered Nb$_3$Sn and V$_3$Si thin films for superconducting radio-frequency cavities}

\author{Katrina Howard$^{1,2}$, Zeming Sun$^1$, and Matthias U. Liepe$^1$}

\address{$^1$ Cornell Laboratory for Accelerator-Based Sciences and Education, Cornell University, Ithaca, NY 14853, USA}
\address{$^2$ Department of Physics, University of Chicago, Chicago, IL 60637, USA}
\ead{khoward99@uchicago.edu (K.H.); zs253@cornell.edu (Z.S.); mul2@cornell.edu (M.U.L.)}
\vspace{10pt}
\begin{indented}
\item[]December 2022
\end{indented}

\begin{abstract}
Nb$_3$Sn and V$_3$Si thin films are promising candidates as thin films for the next generation of superconducting radio-frequency (SRF) cavities. However, sputtered films often suffer from stoichiometry and strain issues during deposition and post annealing. In this study, we explore the structural and chemical effects of thermal annealing, both $\textit{in-situ}$ and post-sputtering, on DC-sputtered Nb$_3$Sn and V$_3$Si films of varying thickness on Nb or Cu substrates, extending from our initial studies \cite{SunRef1}. Through annealing at 950 \textdegree C, we successfully enabled recrystallization of 100 nm thin Nb$_3$Sn films on Nb substrate with stoichiometric and strain-free grains. For 2 $\mu$m thick films, we observed the removal of strain and a slight increase in grain size with increasing temperature. Annealing enabled a phase transformation from unstable to stable structure on V$_3$Si films, while we observed significant Sn loss in 2 $\mu$m thick Nb$_3$Sn films after high temperature anneals. We observed similar Sn and Si loss on films atop Cu substrates during annealing, likely due to Cu-Sn and Cu-Si phase generation and subsequent Sn and Si evaporation. These results encourage us to refine our process to obtain high-quality sputtered films for SRF use.
\end{abstract}

%
\vspace{2pc}
\submitto{Superconductor Science and Technology}
%
\maketitle
%
\ioptwocol

\section{Introduction}

Nb$_3$Sn and V$_3$Si thin films are of increasing interest to the superconducting radio-frequency community owing to the quest of achieving high accelerating gradient and efficiency. As niobium-based superconducting radio-frequency (SRF) cavities are reaching the theoretical limits \cite{SunRef48}, alternative materials are of great interest to continue the quest of increasing quality factors, accelerating gradients, and efficiency \cite{SunRef3}. A15 superconductors Nb$_3$Sn and V$_3$Si are promising candidates for this role, used as thin films inside either Nb or Cu cavities \cite{SunRef3,SunRef2}. Both candidates have relatively high critical temperatures (T$_{c,Nb_3Sn}$ = 18.3 K and T$_{c,V_3Si}$ = 17.1 K), and Nb$_3$Sn is predicted to yield a superheating field of $\sim$ 440 mT that doubles the Nb limit of $\sim$ 240 mT \cite{SunRef3,SunRef4,SunRef5,SunRef6,SunRef7}. These properties could allow cavity operation at an elevated temperature of $\sim$ 4 K and the potential for increased accelerating gradients \cite{SunRef9}. This higher operating temperature allows for reduced cryogenic costs and simpler infrastructure for particle accelerators and their applications \cite{SunRef3}. Due to their brittle nature and low thermal conductivity, Nb$_3$Sn and V$_3$Si are best suited for use as a thin film inside a host cavity with better thermal conductivity, such as Nb or Cu \cite{SunRef3,SunRef10,SunRef11}.

Nb$_3$Sn thin films have been achieved through vapor diffusion, sputtering, electroplating, and chemical vapor deposition \cite{SunRef12,SunRef13,SunRef14,SunRef25,SunRef24,SunRef16}. In the state-of-the-art vapor diffusion, a niobium cavity is placed in a high-temperature vacuum furnace, and then tin or tin chloride sources are vaporized and allowed to diffuse into the niobium surface for alloying \cite{SunRef3,SunRef9,SunRef13,SunRef18,SunRef19,SunRef20}. In contrast, sputtering utilizes high-energy plasma to directly eject target materials onto a substrate at low temperatures \cite{SunRef2,SunRef5,SunRef12,SunRef21,SunRef22}. Alternatively, Nb$_3$Sn films are fabricated via electroplating in aqueous solutions working at near-room temperatures and atmospheric pressure followed by heat treatment \cite{SunRef14,SunRef25,SunRef24,SunRef23}, or via chemical vapor deposition that takes advantage of reactions between volatile precursors \cite{SunRef26}. Nb$_3$Sn has been successfully vapor-diffused inside cavities, where a single-cell reached gradients of 24 MV/m, while Nb$_3$Sn 9- and 5-cells reached 15 MV/m, both with Q$_0$’s on the order of 10$^{10}$ at operating temperature 4.4 K \cite{SunRef19,SunRef20}. In cavity tests, maximum surface fields of 120 mT (pulsed operation) and 80 -- 100 mT (CW) have been achieved, showing that Nb$_3$Sn cavities can be operated reliably in a flux-free metastable state above the lower critical field of this material (around 40 mT) \cite{SunRef6,SunRef27}.

In the sputtering process, the film properties are tailored by controlling the Ar/Kr plasma pressure, substrate temperature, sputtering voltage, sputtering current, rate of deposition, and post-sputtering anneal temperature/duration. In literature \cite{SunRef2,SunRef5,SunRef9,SunRef12,SunRef28}, sputtered Nb$_3$Sn films have been demonstrated on Nb and Cu surfaces by using a stoichiometric Nb$_3$Sn target, by co-sputtering with Nb and Sn targets, or through annealing a sputtered Nb/Sn multilayer. A stoichiometric target allows for a design where only a single target is used \cite{SunRef2,SunRef5,SunRef28,SunRef29}. Co-sputtering involves the use of separate Nb and Sn targets that are sputtering at the same time, allowing for tuning of the power applied to each target \cite{SunRef21,SunRef30}. Multilayer sputtering also uses separate Nb and Sn targets but alternates the use of each target to create many ultrathin layers of each material \cite{SunRef11,SunRef12,SunRef22}. T$_c$’s above 17.8 K have been observed for single-target and multilayer sputtering \cite{SunRef5,SunRef11,SunRef12}.

V$_3$Si films have been attempted by thermal diffusion, magnetron sputtering, and high-power impulse magnetron sputtering (HiPIMS) \cite{SunRef10,SunRef11,SunRef29,SunRef30}. In thermal diffusion, a vanadium layer on a silicon-on-insulator substrate is annealed at high temperature to produce V$_3$Si \cite{SunRef30}. In the HiPIMS method, power is applied as a set of discrete high-energy pulses at a low-duty cycle, which can be used to ion bombard the substrate, recrystallizing films at a low temperature and allowing more control of the stoichiometry; this method of depositing V$_3$Si films on Cu substrates produced T$_c$ up to 10 K \cite{SunRef10}. CERN’s magnetron sputtered V$_3$Si films on a silver buffer layer upon a Cu substrate have reached T$_c$ of 11.2K \cite{SunRef29}.

Thermal annealing of the sputtered films, either $\textit{in situ}$ or post-deposition, is required to minimize the internal stress induced by the sputtering process and improve the stoichiometry and grain structures, which are important for their critical temperature and cavity RF performance \cite{SunRef2,SunRef5,SunRef12}. However, during annealing of sputtered Nb$_3$Sn or Nb/Sn multilayers, the films suffer from issues such as Sn loss, Cu incorporation into the film from Cu substrates, high strain, and interface issues at the substrate-film boundary \cite{SunRef2,SunRef5,SunRef12}. Sn loss is a critical issue because of the dependence of T$_c$ on Sn concentration \cite{SunRef3}. While annealing is frequently performed on Nb$_3$Sn films, these high temperatures have led to Sn loss in the furnace and Nb-rich films with reduced T$_c$ \cite{SunRef5,SunRef12}, which motivates us to mechanistically understand the phase transformation associated with annealing. Cu incorporation can occur during annealing, which lowers the T$_c$ \cite{SunRef2}. This issue can be addressed by using a barrier layer such as tantalum to reduce the interdiffusion \cite{SunRef29}. The interface between Nb$_3$Sn and Cu also suffers from strain because of their different thermal expansion coefficients and lattice mismatch, which can cause cracking in the film \cite{SunRef2}. Cracking can release high initial strain in the lattice, but does not relieve microstrain and increases surface roughness while decreasing the uniformity of the film \cite{SunRef2,SunRef31}. Currently, no sputtered Nb$_3$Sn cavity test has been reported. Moreover, V$_3$Si is much less studied than Nb$_3$Sn, and there has been no RF test to date \cite{SunRef10,SunRef11,SunRef29}.

One goal of this work is to optimize the sputtering capability of these alternative SRF materials at Cornell and compare our results with existing efforts in the SRF field. Most importantly, we aim to systematically investigate the effect of thermal annealing on the sputtered Nb$_3$Sn and V$_3$Si thin films in order to better understand these observed issues and design an optimal process for SRF use. By understanding the impacts of deposition and annealing parameters, our goal is to find the root of the issues in stoichiometry and strain of thin films. With such knowledge, we hope to provide insights for the development of sputtered Nb$_3$Sn and V$_3$Si cavities. In this study, we investigate Nb$_3$Sn and V$_3$Si films of different thicknesses on both Nb and Cu substrates to optimize the best conditions that minimize strain while producing required stoichiometry and superconducting properties.

\begin{table*}[htbp]
\caption{Sputtering parameters for Nb$_3$Sn and V$_3$Si film deposition.}
\begin{tabular*}{\textwidth}{p{1.5cm} p{1.5cm} p{2cm} p{2cm} p{2cm} p{2cm} p{2cm}}
\br
Film & Substrate	& Substrate holder	& Temperature (\textdegree C) & Voltage (V) & Current (A)	& Nominal thickness \\
\mr
Nb$_3$Sn	& Nb	& Rotating	& 25	& 596	& 0.15	& 100 nm\\
Nb$_3$Sn	& Nb	& Rotating	& $>$ 25 &	589	& 0.26	& 2 $\mu$m\\
Nb$_3$Sn	& Cu	& Heated	& 550	& 466	& 0.214	& 300 nm\\
V$_3$Si	& Nb	& Rotating	& $>$ 25 &	811	& 0.196	& 2 $\mu$m\\
V$_3$Si	& Cu	& Heated	& 550	& 819	& 0.222	& 300 nm\\
\br
\end{tabular*}
\label{T1}
\end{table*}

\section{Methods}

Nb$_3$Sn and V$_3$Si thin films were deposited using a DC-sputtering system at the Cornell Center for Materials Research. A high vacuum of 10$^{-6}$ torr base pressure was achieved using a cryo-pumped system. All depositions were performed at 5 mTorr Ar pressure. A rotating stage was used, when possible, to ensure uniformity during deposition.

As summarized in Table~\ref{T1}, the sputtering parameters varied were the film material (Nb$_3$Sn vs. V$_3$Si), substrate material (Nb vs. Cu), deposition temperature (room temperature vs. 550 \textdegree C $\textit{in situ}$ heating), and film thickness (100 nm, 300 nm, and 2 $\mu$m). Bulk Nb$_3$Sn and V$_3$Si targets were used, and they were purchased from ACI alloy, Inc. The impurity concentrations as received were 0.01 at.\%. 

Nb and Cu squared substrates of 1 cm$^2$ area were used in order to provide insights for applications in Nb and Cu substrate cavities. Before deposition, Nb substrates were electropolished, and Cu substrates were chemically polished to ensure a smooth surface.

The Nb$_3$Sn and V$_3$Si films were designed to have thicknesses of 100 nm and 2 $\mu$m on Nb substrates and 300 nm on Cu substrates. The deposition rate was 2.5 Å/s for all samples except for the V$_3$Si film on Cu substrate which was 1.8 Å/s (as there was difficulty lighting the plasma). The deposition temperature for the thick 2 $\mu$m samples is subject to error because the temperature is uncontrolled upon the rotating stage and increased through the 133-minute deposition. Subsequently, a 550 ℃ heating stage was applied to investigate the effect of $\textit{in situ}$ heating during deposition.

After the sputtering process, films were annealed in a series of elevated temperatures at 600 ℃, 700 ℃, 800 ℃, and 950 ℃, each for 6 hours, in a Lindberg high-vacuum (3 $\times$ 10$^{-7}$ Torr) furnace. The heating rate was 10 ℃ per minute, and the annealing was followed by furnace cooling.

Structural and chemical analyses were conducted between anneals to characterize the films. These analysis methods included scanning electron microscope (SEM) to observe the grain structure and size, energy dispersive X-ray (EDS) and X-ray photoelectron (XPS) spectroscopies to determine the atomic composition, and X-ray diffraction (XRD) to gain insight into the crystal structure of the film and calculate the strain. In this analysis, the key features are the quality of the film surfaces (smoothness, uniformity, grain shape/size), the stoichiometry of the films, and the existence and strain of Nb$_3$Sn and V$_3$Si diffraction planes. Note that EDS results were calibrated with regard to the electron penetration depth in each material and the film thickness.

Finally, on the 100 nm thin Nb$_3$Sn film that yields the best performance, we verified its critical temperature using a quantum design physical property measurement system (PPMS) and quantified the surface roughness using atomic force microscopy (AFM).

\section{Results and Discussion}

In this section, we first analyze the recrystallization behavior observed in the 100 nm thin Nb$_3$Sn films annealed and discuss the superconducting, composition, and surface properties of these films. Next, we show the composition and strain evolutions as a function of annealing temperature in the 2 $\mu$m thick Nb$_3$Sn and V$_3$Si films (on Nb) and attempt to understand the Sn loss and strain relief mechanisms. Finally, we show the ternary phase transformation upon annealing in the 300 nm thick Nb$_3$Sn and V$_3$Si films that were deposited on Cu substrates. Representative surface morphologies of samples upon deposition and after 700 \textdegree C and 950 \textdegree C anneals are shown in figure S1.

\subsection{Thin Nb$_3$Sn film: Recrystallization}

\subsubsection{Recrystallization behavior}

\begin{figure}[htbp]
\centering
\includegraphics[width=\linewidth]{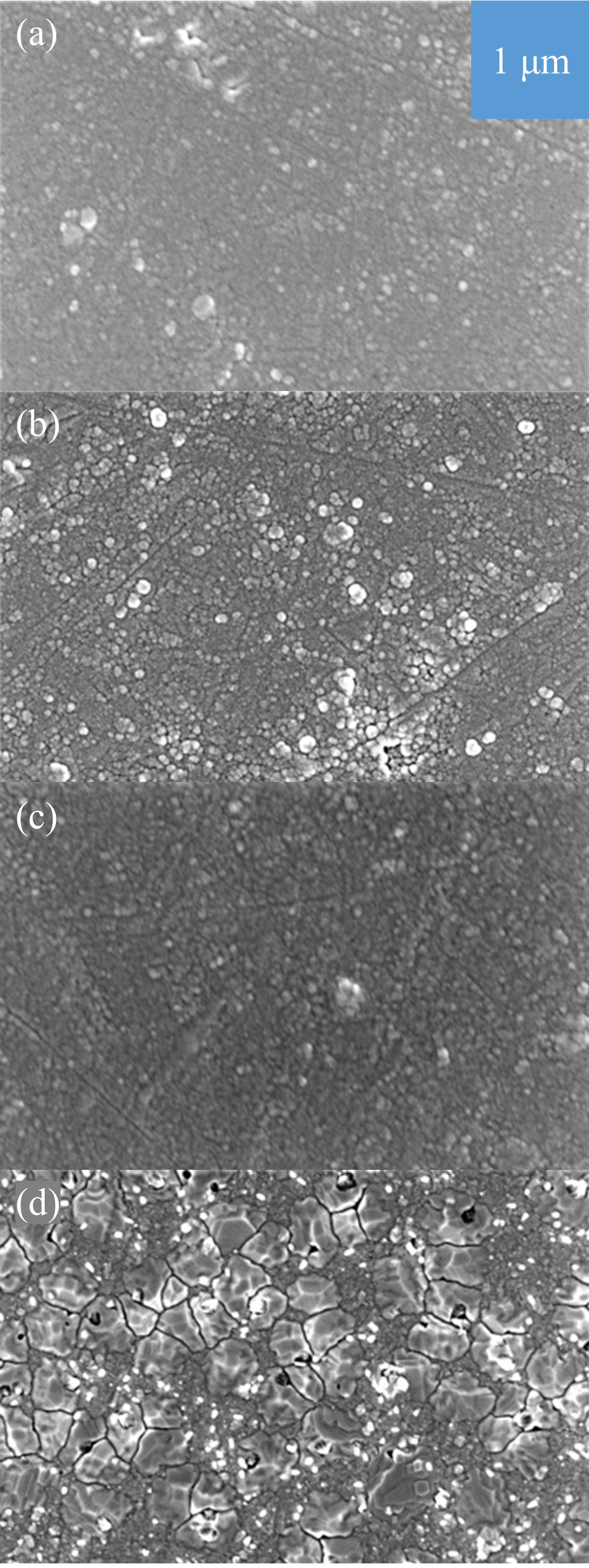}
\caption{Surface SEM images for 100 nm thin Nb$_3$Sn films on Nb substrates: (a) as-deposited (a), and (b-d) after annealing: (b) 600 ℃, (c) 800 ℃, and (d) 950 ℃.}
\label{SunFig1}
\end{figure}

\begin{figure*}[htbp]
\centering
\includegraphics[width=\linewidth]{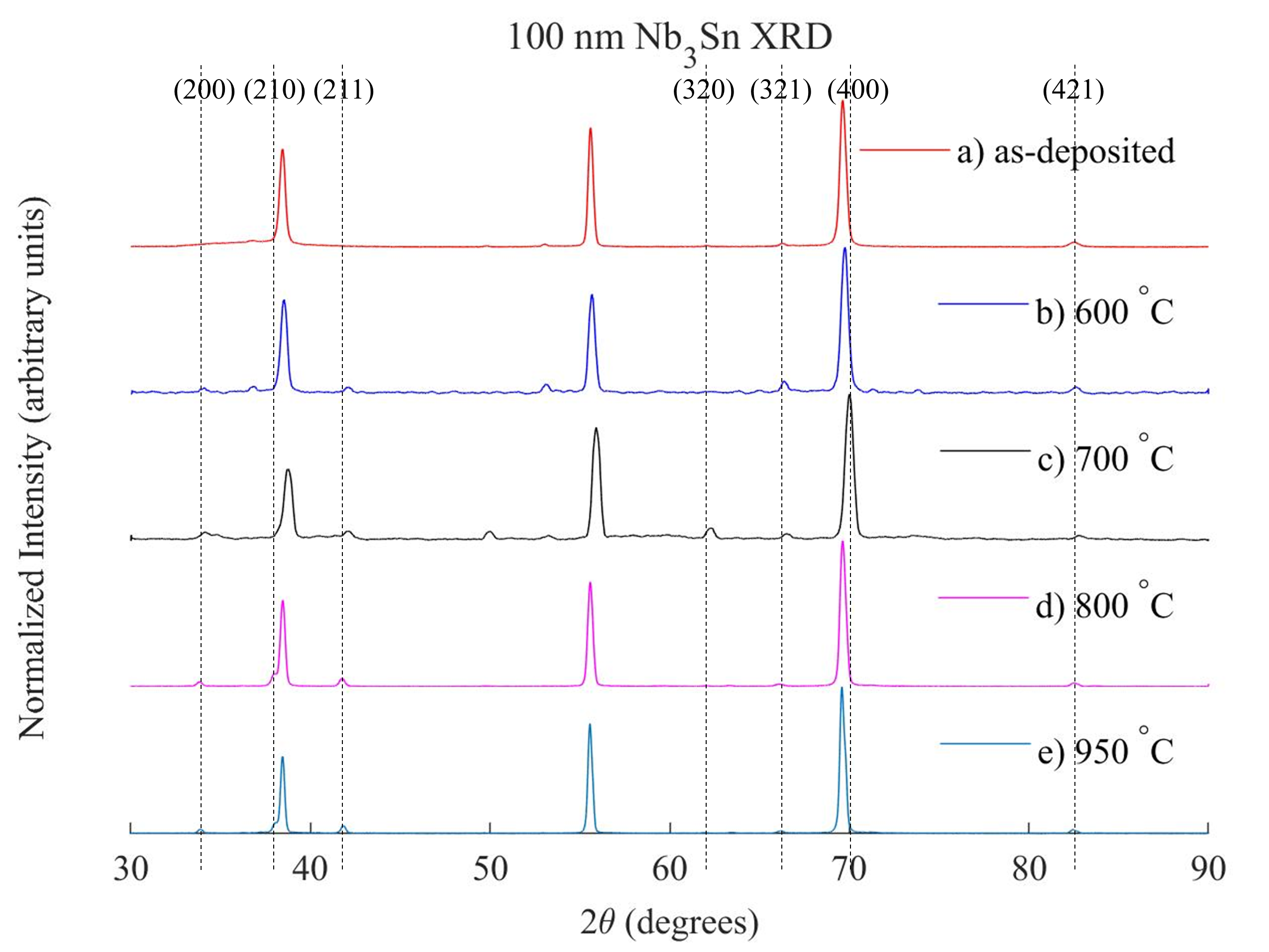}
\caption{XRD patterns taken from the 100 nm thin Nb$_3$Sn film as a function of annealing temperature: (a) as-deposited, (b) 600 ℃, (c) 700 ℃, (d) 800 ℃, (e) 950 ℃. Observed Nb$_3$Sn diffraction planes are labeled at the top.}
\label{SunFig2}
\end{figure*}

Figure~\ref{SunFig1} shows the evolution of surface morphology with increasing temperature for the 100 nm thin Nb$_3$Sn film on Nb substrate. We observed evident grain recrystallization at 950 ℃ anneals. The grain size increased from a few nanometers as deposited (figure~\ref{SunFig1}a) to approximately 300 nm after annealing (figure~\ref{SunFig1}d). Recrystallization occurs through the release of strain energy during annealing and the subsequent migration of grain boundaries \cite{SunRef32}.

Here, we discuss the driving force and boundary mobility for thermodynamic considerations of this recrystallization annealing. The stored energy per unit volume (E$_s$) at a strain level ($\epsilon$) of 0.2, the maximum strain measured from our sputtered films, is 2.7 × 10$^{9}$ J/m$^3$, based on a 1-dimensional elastic assumption, E$_s$$\,$=$\,$1/2$\,$E$\epsilon^2$, where E is Young’s modulus and the value for Nb$_3$Sn at 300 K is 13.7 × 10$^{11}$ dyn/cm$^2$ \cite{SunRef33}. Indeed, this value is dramatically larger than the typical lightly-deformed energy of 10$^5$ J/m$^3$ for driving recrystallization in metals \cite{SunRef34}. This suggests a sufficient driving force from the sputtering-induced strain within the film to enable the recrystallization annealing.

Our X-ray diffraction data (figure~\ref{SunFig2}) shows the Nb$_3$Sn phase is consistent in terms of grain orientation at all annealing temperatures including 950 ℃ recrystallizations. We find Nb$_3$Sn peaks near the known powder diffraction peaks at 2$\theta$ = 33.6\textdegree, 37.7\textdegree, 41.5\textdegree, 62.8\textdegree, 65.6\textdegree, 70.6\textdegree, and 82.9\textdegree \cite{SunRef36}. Due to the large penetration depth of the X-ray probe, strong Nb substrate diffractions are seen at 2$\theta$ = 38.4\textdegree, 53.3\textdegree, and 69.3\textdegree. A complete list of known peak locations is shown in table S1. As the annealing temperature was increased to 950 ℃, the grain orientations of Nb$_3$Sn remained while the growth of grain size was significant.

We assume this recrystallization follows a boundary migration mechanism and evaluate the boundary mobility by the Arrhenius law, d$\,$=$\,$A$\,$×$\,$exp$\,$(-$\,$E$_a$$\,$/$\,$RT), where d is the equilibrium grain size, A is the pre-exponential factor, E$_a$ is the activation energy, and T is annealing temperature. The apparent values of the pre-exponential factor and activation energy were determined to be 2.59 × 10$^5$ and 63 ± 2 kJ/mol, respectively, by Schelb \cite{SunRef35}. At an annealing temperature of 950 ℃, the maximum attainable grain size is in the range of 434 -- 643 nm. The observed $\sim$ 300 nm grain size in our work is reasonable considering the influence from annealing time (6 h in our work versus up to 200 h in Schelb’s work).

In summary, recrystallization anneal above 800 ℃ is effective in relieving the built-in strain from sputtering and thus forming stoichiometric Nb$_3$Sn along with grain coarsening. 

\subsubsection{Film properties}

\begin{figure}[htbp]
\centering
\includegraphics[width=\linewidth]{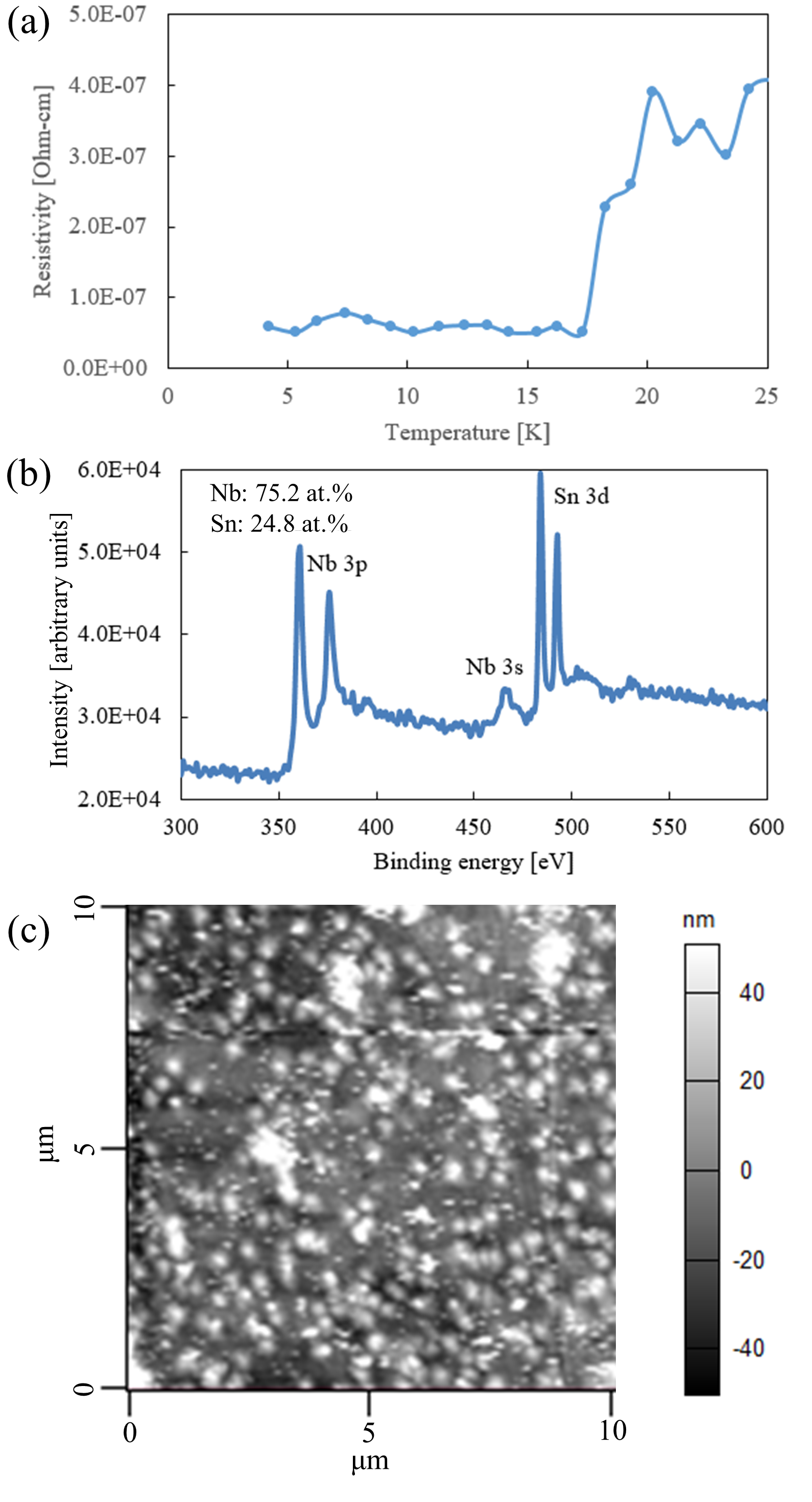}
\caption{Film properties for the 100 nm thin Nb$_3$Sn film on a Nb substrate after 950 ℃ annealing. (a) Resistive transition and critical temperature, (b) XPS spectrum showing the atomic composition after sputtering away the surface 20 nm layer, and (c) AFM image showing low surface roughness.}
\label{SunFig3}
\end{figure}

Superconducting properties, atomic composition, and surface roughness were investigated on the 100 nm Nb$_3$Sn sample after the 950 ℃ annealing for 6 hours. As shown in figure~\ref{SunFig3}, the critical temperature is determined to be 17.5 K, while the Nb/Sn stoichiometry is 3/1 after sputtering away the surface oxides. Similarly, Sayeed et al. \cite{SunRef5} reported T$_c$ values of 17.68 -- 17.83 K for 350 nm Nb$_3$Sn sputtered films that were annealed at 800 ℃ for 24 hours and 1000 ℃ for 1 hour with low Sn loss. They observed significant degradation of T$_c$ down to 10.95 K as a consequence of the dramatic Sn loss down to 4\% after annealing for 24 h at 1000 ℃. In contrast, we did not observe the Sn loss in the 100 nm thin films after annealing. We infer recrystallization plays a major role in retaining the Sn ratio as well as maintaining T$_c$ $\sim$ 17.5 K in the 100 nm thin films, with the relatively short annealing time helping to retain the film properties. However, our 2 $\mu$m thick films as detailed in Section 3.2 showed similar Sn loss behavior at increasing annealing temperatures as compared to the 350 nm thick films in Sayeed’s work \cite{SunRef5}, which indicates the importance of a recrystallization process to obtain stoichiometric Nb$_3$Sn films with T$_c$ 17.5 K.
Additionally, the atomic force microscopy (AFM) result is shown in figure 3c. The film shows low surface roughness, with an average roughness of 18.3 nm, RMS roughness of 25.3 nm, and a maximum height difference of 600 nm. In contrast, Nb substrates used have an average roughness of $\sim$ 70 nm.

\subsection{Thick Nb$_3$Sn and V$_3$Si films: relation of strain and composition change}

\begin{figure}[htbp]
\centering
\includegraphics[width=\linewidth]{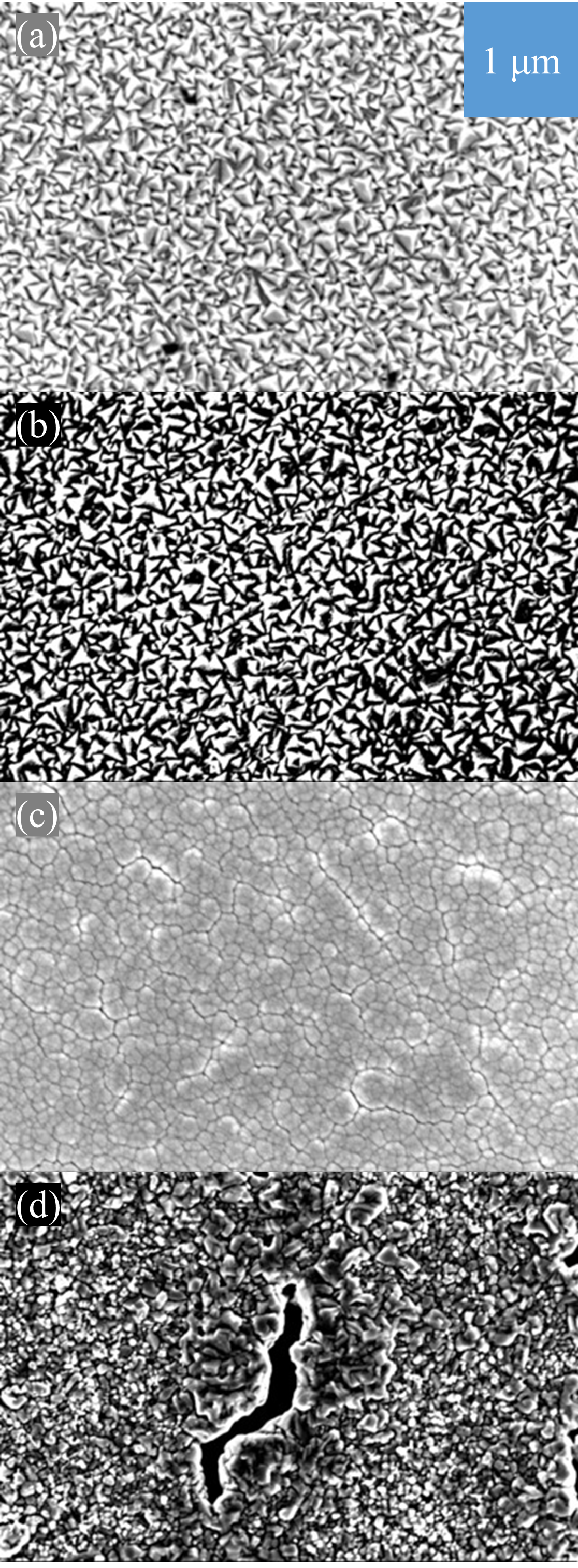}
\caption{Surface SEM images for 2 $\mu$m thick Nb$_3$Sn (a, b) and V$_3$Si (c, d) films on Nb substrates: (a, c) as-deposited and (b, d) after 950 ℃ annealing.}
\label{SunFig4}
\end{figure}

Thermal annealing was performed on 2 $\mu$m thick Nb$_3$Sn and V$_3$Si films on the Nb substrate. The initial thickness of the films greatly altered the annealing behaviors as compared to results from 100 nm thin films.

\subsubsection{2 $\mu$m thick Nb$_3$Sn films}

The Nb$_3$Sn grains nucleated in a triangular shape and remained in that shape at all annealing temperatures studied (figure~\ref{SunFig4}a~and~\ref{SunFig4}b). We speculate that these small triangular-shaped grains with 100 -- 200 nm in size were induced by the high built-in stress and subsequent plastic deformation during deposition. In-plane stress is typical in physical vapor sputtering, and small grains with angular shapes are favored under the stress \cite{SunRef37}. This argument is supported by the $\textit{in situ}$ stress versus grain size relationship in Leib’s work \cite{SunRef38}.

\begin{figure}[htbp]
\centering
\includegraphics[width=\linewidth]{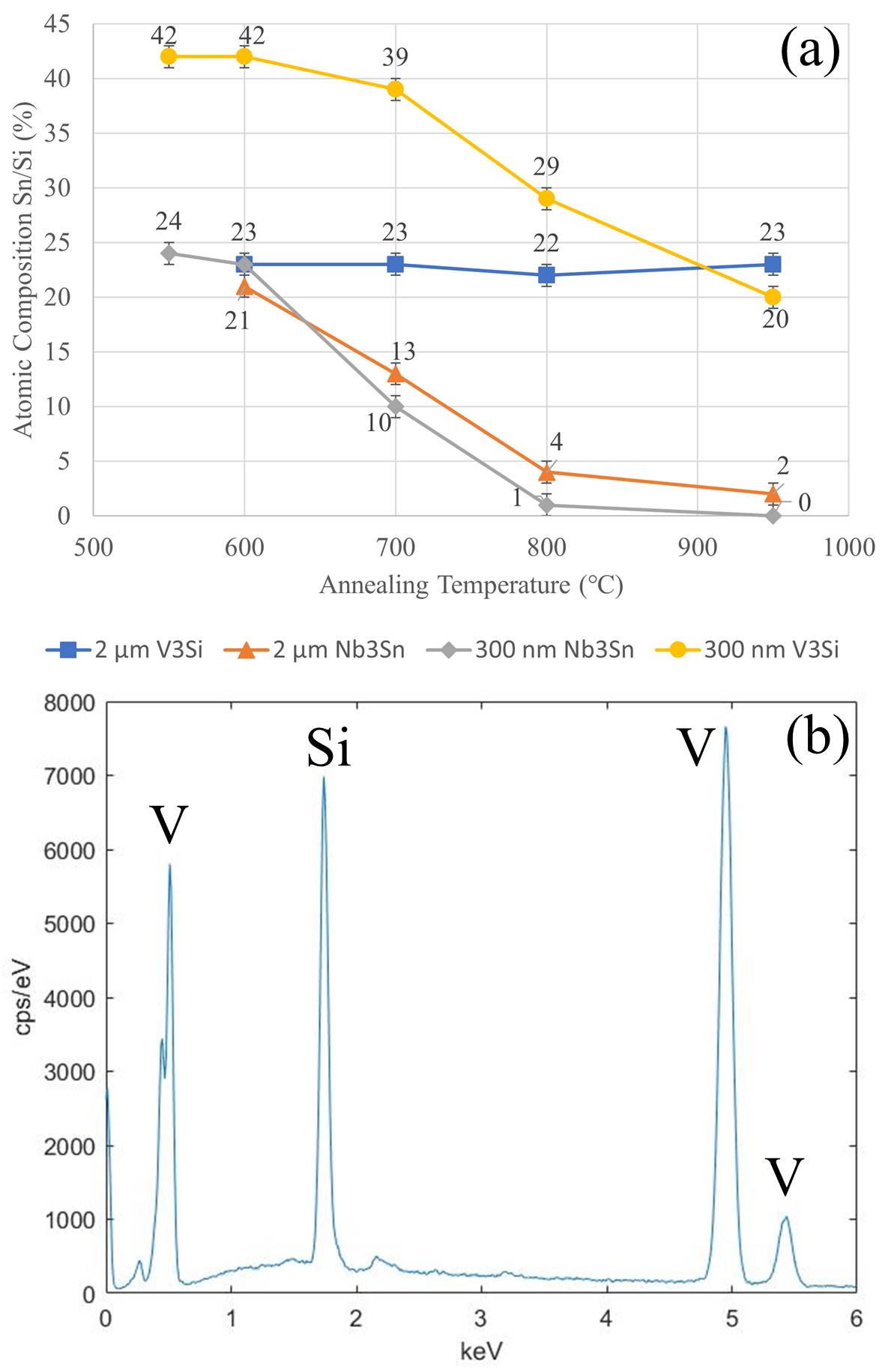}
\caption{(a) Sn/[Sn+Nb] or Si/[Si+V] ratios in the Nb$_3$Sn and V$_3$Si films, respectively, as a function of annealing temperature for the 2 $\mu$m thick films sputtered on Nb substrates and the 300 nm thick films on Cu substrates (discussed in Section 3.3). Note that the high Si ratios for Cu substrate samples are due to exclusion of Cu signals for calculation. As-deposited Sn/Si composition on 2 $\mu$m thick films are 23 -- 25 \%. (b) Example of the EDS spectrum taken on the 2 $\mu$m thick V$_3$Si film for generating the composition dataset.}
\label{SunFig5}
\end{figure}

\begin{figure}[htbp]
\centering
\includegraphics[width=\linewidth]{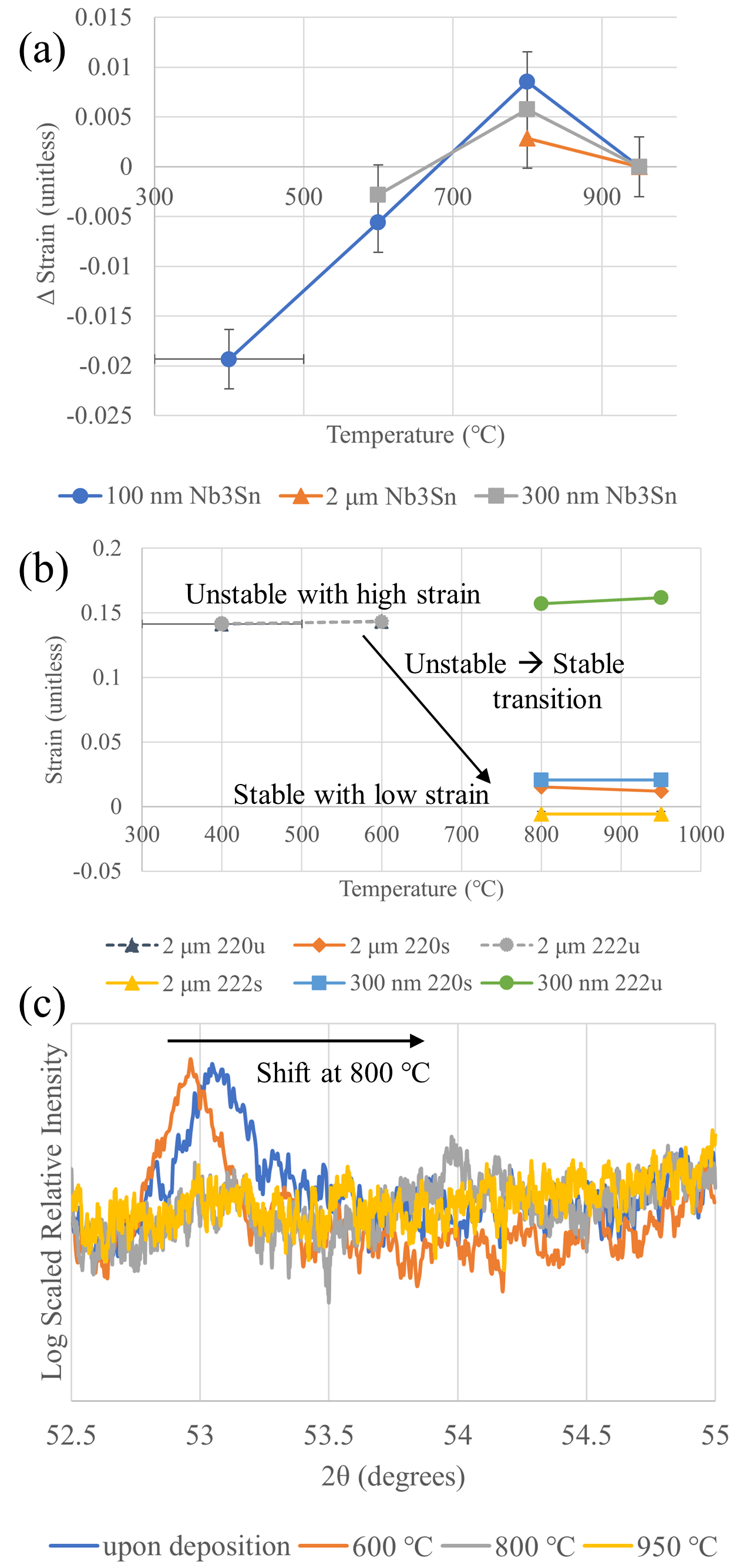}
\caption{(a) Relative strain that is normalized to the high-temperature anneal limit, as a function of annealing temperature for the 100 nm and 2 $\mu$m Nb$_3$Sn films on Nb substrates together with the 300 nm Nb$_3$Sn film on Cu substrates. (b) Temperature-dependent strain diagram calculated from stable (s) and unstable (u) (220) diffraction peak shifting for 2 $\mu$m and 300 nm V$_3$Si films on the Nb and Cu substrates, respectively. (c) Example of the XRD patterns taken on the 2 $\mu$m thick V$_3$Si film showing the stable and unstable (220) diffraction peaks for generating the strain diagram.}
\label{SunFig6}
\end{figure}

Upon annealing, as shown in figure~\ref{SunFig5}a, the 2 $\mu$m thick Nb$_3$Sn films experienced significant Sn loss from the as-deposited $\sim$ 24\% down to 21\% after the initial anneal at 600 ℃ and further down to nearly 2\% after the 950 ℃ anneal. Conversely, Nb$_3$Sn phases were barely observed in the X-ray diffraction until Nb$_3$Sn peaks appeared at 800 ℃ and 950 ℃ anneals. The strain ($\epsilon$) for a given plane was calculated by $\epsilon$$\,$=$\,$(a$_T$$\,$-$\,$a$_0$)$\,$/$\,$a$_0$, where a$_T$ is the measured lattice constant from Nb$_3$Sn plane diffraction and a$_0$ is the lattice parameter from database \cite{SunRef36}. (Internal strains calculated for all samples are summarized in Table S2.) The relative strain ($\Delta\epsilon$), shown in figure~\ref{SunFig6}a, was obtained by normalizing strain to the high-temperature anneal limit where we observe negligible strains. 

Here, we analyze the effect of film thickness on the strain. The internal strain ($\epsilon$) in a biaxial thin film system where in-plane stresses are equal ($\delta$$\,$=$\,$$\delta_{11}$$\,$=$\,$$\delta_{22}$) can be described by a linear relationship as $\epsilon$$\,$=$\,$(2$\,$S$_{1}$$\,$+$\,$1/2$\,$S$_{2}$$\,$sin$^2\phi$)$\,$$\delta$, where $\phi$ is the angle from the film normal to the diffraction plane normal, and S$_1$ and S$_2$ are the X-ray elastic constants that are determined, in an elastic isotropic scenario, by Young’s modulus (E) and Poisson’s ratio ($\upsilon$) and given by –$\,$$\upsilon$$\,$/$\,$E and (1$\,$+$\,$$\upsilon$)$\,$/$\,$E, respectively \cite{SunRef38}. The built-in stress increases with film thickness (or deposition time at a fixed deposition rate) in a polycrystalline film system when the deposition goes beyond the initial instantaneous stress stage ($<$ 10 nm thickness) \cite{SunRef37}. This positive correlation, although slightly affected by the growth-interrupt stress relaxation effect and the heating effect, suggests high strain in the 2 µm thick film; however, the high in-plane stress during thicker film sputtering results in plastic deformation as indicated by the observation of small grain sizes and high density of boundaries (figure~\ref{SunFig4}a).

Furthermore, we cannot fully explain the Sn loss in the course of annealing the sputtered Nb$_3$Sn films, $\textit{i.e.}$, the decrease of Sn/(Nb+Sn) atomic ratios with the increasing annealing temperature (figure~\ref{SunFig5}a). Note that this Sn loss behavior is repeatedly observed in previous Nb$_3$Sn sputtering work \cite{SunRef5,SunRef12}. At the annealing temperatures studied, pure Sn phases are not expected due to their low vaporization temperatures ($\textit{e.g.}$, 800 ℃ at 10$^{-6}$ Torr), so we primarily consider Nb-Sn alloy phases in the film. The as-deposited film showed a 23 -- 25\% Sn atomic ratio which suggests minimal Sn-rich phases (Nb$_6$Sn$_5$ and NbSn$_2$) based on the Nb-Sn phase diagram \cite{SunRef3}; these Sn-rich phases were also not observed in the X-ray diffraction. Without Sn or Sn-rich phases, merely Nb$_3$Sn is expected in this study as indicated by the 23 -- 25\% Sn ratio, but XRD did not show any detectable Nb$_3$Sn diffractions upon deposition; the crystalline Nb$_3$Sn phase has an extremely high ($>$ 2100 ℃) phase transformation temperature, making it unlikely to explain the Sn loss. We, therefore, suspect the generation of amorphous Nb$_3$Sn phases in the film. Such amorphous phases were reported when using non-equilibrium processing techniques \cite{SunRef40,SunRef41}. This could cause the loss of Sn alloys via the generation of $\alpha$-Nb and also explain the appearance of Nb$_3$Sn diffraction for anneals above 800 ℃, which likely corresponds to the crystallization temperature. This requires further investigation.

\subsubsection{2 $\mu$m thick V$_3$Si films}

Different from thick Nb$_3$Sn films, the as-deposited 2 $\mu$m thick V$_3$Si film (figure~\ref{SunFig6}b) exhibits a high strain of 15\%, which supports the positive relationship between strain and film thickness in an elastic scenario for thick ($>$ 10 nm) polycrystalline films. The initial film shows a near-stoichiometric value of Si ($\sim$ 23\%) shown in figure~\ref{SunFig5}b. 

Upon annealing, the V$_3$Si film shows a constant Si concentration for all temperatures; see figure~\ref{SunFig5}a. In contrast, the strain within the film is significantly relieved together with a transition from the unstable V$_3$Si structure to the stable structure between 800 ℃ and 950 ℃ (figure~\ref{SunFig6}b). The structural transformation is observed through the shifting of the (220) and (222) diffraction peaks in figure~\ref{SunFig6}c. These behaviors demonstrate that thermal annealing contributes to strain reduction and structural stabilization in a thick sputtered film.
However, as shown in figure~\ref{SunFig4}d, large cracks begin to appear on the film after the first anneal at 600 ℃, coinciding with a shift toward a more angular grain shape with increasing temperature. The high strain induced by the sputtering deposition is responsible for the cracks although thermal relaxation has reduced a significant amount of lattice strain.  

\subsection{Nb$_3$Sn and V$_3$Si films on Cu substrates: ternary alloy systems}

\begin{figure}[htbp]
\centering
\includegraphics[width=\linewidth]{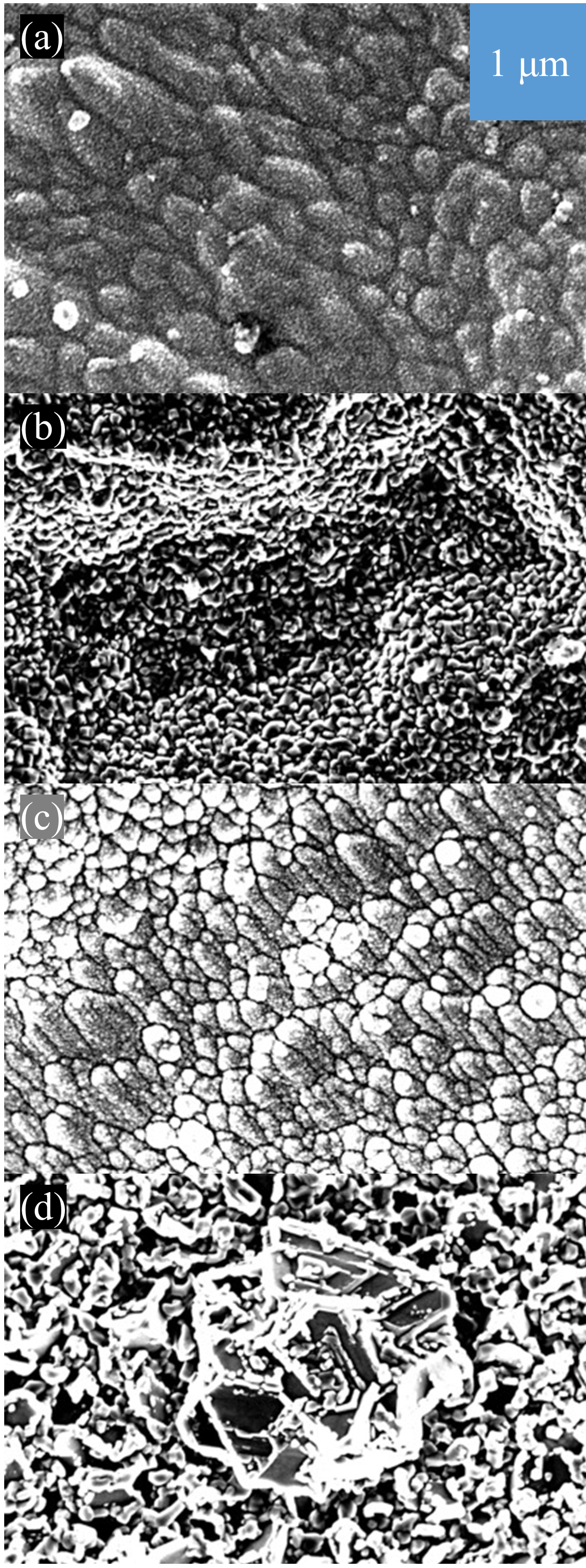}
\caption{Surface SEM images for 300 nm Nb$_3$Sn (a, b) and V$_3$Si (c, d) films on Cu substrates: (a, c) as-deposited and (b, d) after 950 ℃ annealing.}
\label{SunFig7}
\end{figure}

By studying the temperature-atomic percentage phase diagrams of Nb-Sn \cite{SunRef3} and V-Si \cite{SunRef47}, as well as the three-element composition phase diagrams of Cu-Nb-Sn \cite{SunRef42,SunRef43,SunRef44} and Cu-V-Si \cite{SunRef45,SunRef46}, we can gain insight into the phase transformations our films undergo during the annealing process. 

As shown in figures~\ref{SunFig7}a~and~\ref{SunFig7}c, 300 nm thick Nb$_3$Sn and V$_3$Si films were sputtered on Cu substrates using a 550 ℃ $\textit{in situ}$ heating stage.  Upon annealing, both films undergo dramatic grain structure changes due to the generation of Cu-Sn or Cu-Si phases. Nb$_3$Sn grains start with rounded grains collecting in finger-like formations as deposited on a Cu surface (figure~\ref{SunFig7}a) and they remelt into small angular grains collecting in regions of differing densities after 950 ℃ anneal (figure~\ref{SunFig7}b). In contrast, V$_3$Si grains begin with a finger-like pattern after deposition (figure~\ref{SunFig7}c) and end with small angular grains and large artifacts scattered across the surface after 950 ℃ anneal (figure~\ref{SunFig7}d). Overall, there is a trend of grain angularization and pattern restructuring with increasing temperature.

The ternary phase transformation that includes Cu-alloy in the films primarily determines the film properties. As shown in figure~\ref{SunFig5}a, the 300 nm Nb$_3$Sn films on Cu substrates suffer from the Sn loss similar to 2 $\mu$m thick films on Nb substrates, but the mechanism is different. According to the Nb-Sn-Cu phase diagram \cite{SunRef42,SunRef43,SunRef44}, Cu-Sn and Nb-Sn phases generate at low temperatures ($\textit{e.g.}$, 675 ℃ \cite{SunRef42}) and these Cu-Sn transform into liquid at 800 ℃ under atmospheric pressure \cite{SunRef43}, and at high temperatures (e.g., 1000 ℃ \cite{SunRef44}), only Nb$_3$Sn and Cu exist. In our study, high-vacuum annealing vaporized the Cu-Sn phases leading to a continuous loss of Sn with increasing temperature. Also, we observed low-intensity Nb$_3$Sn diffraction at all temperatures whereas convoluted XRD peaks that are possibly from Cu-Sn and other Nb-Sn phases appeared at low temperatures. These observations match with the existence of Nb$_3$Sn in the phase diagram at high temperatures although the majority of the film was evaporated.   

Different from Nb-Sn-Cu, the V-Si-Cu phase diagram \cite{SunRef45,SunRef46} shows Cu-Si and V-Si phases at low temperatures ($\textit{e.g.}$, 700 ℃ \cite{SunRef45}), but there is no liquid phase at high temperatures (e.g., 800 ℃ \cite{SunRef46}). Instead, these phases transform into V-Si and Cu phases. In our study, as shown in figure~\ref{SunFig5}a, the Si/(Si+V) ratio begins with a high value of 42\% due to the presence of Cu-Si phases generated during the 550 ℃ $\textit{in situ}$ heated deposition; note that Cu signal is evident, but is not included in the calculation. After annealing, the Si/(Si+V) ratio drops to 20\% at 950 ℃. This phenomenon strongly supports the disappearance of Cu-Si phases in the phase diagram, and only V-Si phases together with some Cu metallic inclusions are expected in the annealed films. Our diffraction data suggest these V-Si phases include the stable (220) and unstable (222) V$_3$Si structures.    

\section{Conclusions and Outlook}

In this study, we have demonstrated the capability of annealing the sputtered thin films to produce successful Nb$_3$Sn and V$_3$Si surfaces that have the potential for use inside SRF cavities. We observe that annealing is required to release the strain in the film and promote grain growth. For our Nb$_3$Sn samples, the best results are found on the recrystallized 100 nm film, where large grains form at 950~℃ anneals. These films are also smooth and have minimal surface defects. The 2 $\mu$m Nb$_3$Sn films are not able to overcome the built-in stress and plastic deformation during sputtering, and likely form an amorphous Nb-Sn phase that leads to nearly complete Sn loss upon annealing. In contrast, the V$_3$Si samples retain the stoichiometry at high temperatures, along with a transition in the grain shape to become more angular. Most interesting was the behavior of these films with respect to the unstable and stable phases of V$_3$Si. In the 2 $\mu$m film, there was a complete transition from unstable to stable at 800 ℃ along with consistent stoichiometry. Because we observe this transition and the proper stoichiometry at high temperatures, we determine these are successful V$_3$Si films.

For the Cu substrate samples, 550 \textdegree C \textit{in situ} heated deposition and the subsequent low-temperature anneals produce Cu-Si and Cu-Sn phases. These phases transform at high temperatures, extracting high concentrations of Cu inclusions in the film. The Cu impurities and Cu-related phases could adversely affect the SRF performance of Nb$_3$Sn/V$_3$Si films inside Cu cavities. In a future study, we would be interested in the use of an ultrathin buffer layer between the Cu and the superconducting layer to prevent this effect \cite{SunRef29}.

In our results, we observed a similar Sn loss as in previous studies \cite{SunRef5}. We are interested in finding ways to prevent this loss such as minimizing strong undercooling and avoiding disordered Nb-Sn phases or using encapsulation during the annealing process. We would like to obtain the benefits of annealing such as recrystallization and strain removal while avoiding events such as Sn loss and cracking. Because the 100 nm Nb$_3$Sn film was successful, it would be important in a future study to further investigate films of similar thickness to optimize grain growth and RF performance.

\section*{Data Availability Statement}

The data that support the findings of this study are available
upon reasonable request from the authors.

\section*{Conflicts of Interest}

The authors declare no competing financial interests.

\section*{Acknowledgements}

This work was supported by the U.S. National Science Foundation under Award PHY-1549132, the Center for Bright Beams. This work made use of the Cornell Center for Materials Research Shared Facilities which are supported through the NSF MRSEC program (DMR-1719875).

\clearpage
\onecolumn
\section*{Appendix}
\makeatletter 
\renewcommand{\thefigure}{S\@arabic\c@figure}
\renewcommand{\thetable}{S\@arabic\c@table}
\makeatother

\setcounter{figure}{0}
\begin{figure*}[hbt!]
    \centering
    \includegraphics[width= \linewidth]{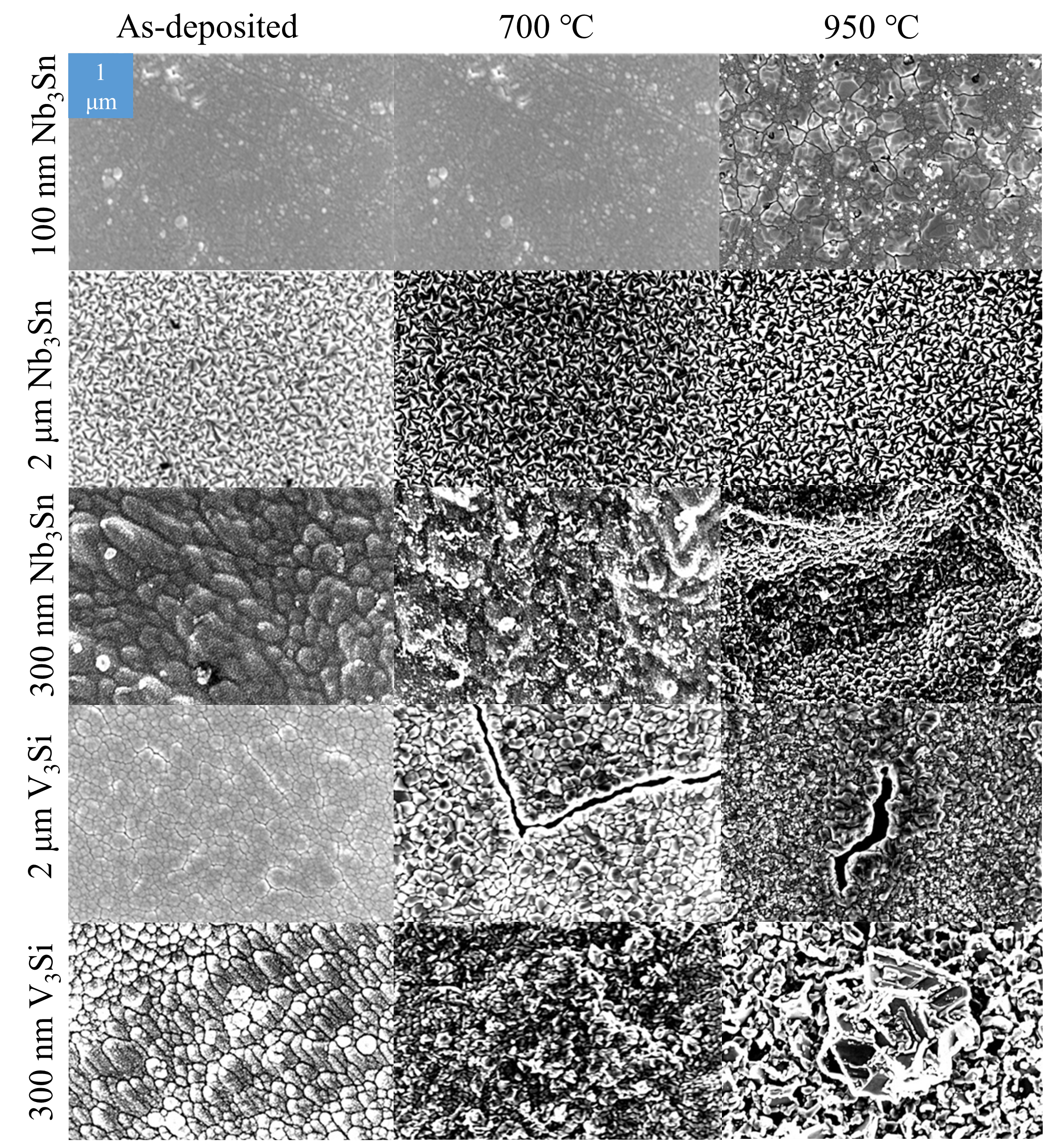}
    \caption{SEM map of all samples upon deposition, after 700 \textdegree C anneal, and after 950 \textdegree C anneal.}
\end{figure*}

\setcounter{figure}{0}
\begin{figure*}[!h]
    \centering
    \renewcommand\figurename{Table}
    \caption{X-ray diffraction (XRD) peaks of Nb$_3$Sn, V$_3$Si (stable vs. unstable), substrates Nb and Cu, and other possibly relevant phases (NbSn$_2$, Nb$_6$Sn$_5$, V$_5$Si$_3$, V$_6$Si$_5$, Nb$_3$Cu, V$_3$Cu  Cu$_{15}$Si$_4$, Cu$_3$Si$_4$, and V (unstable) from reference \cite{SunRef36}. For NbSn$_2$, Nb$_6$Sn$_5$, V$_5$Si$_3$, and V$_6$Si$_5$ peaks, we only listed the prominent points.}
    \includegraphics[width=\linewidth]{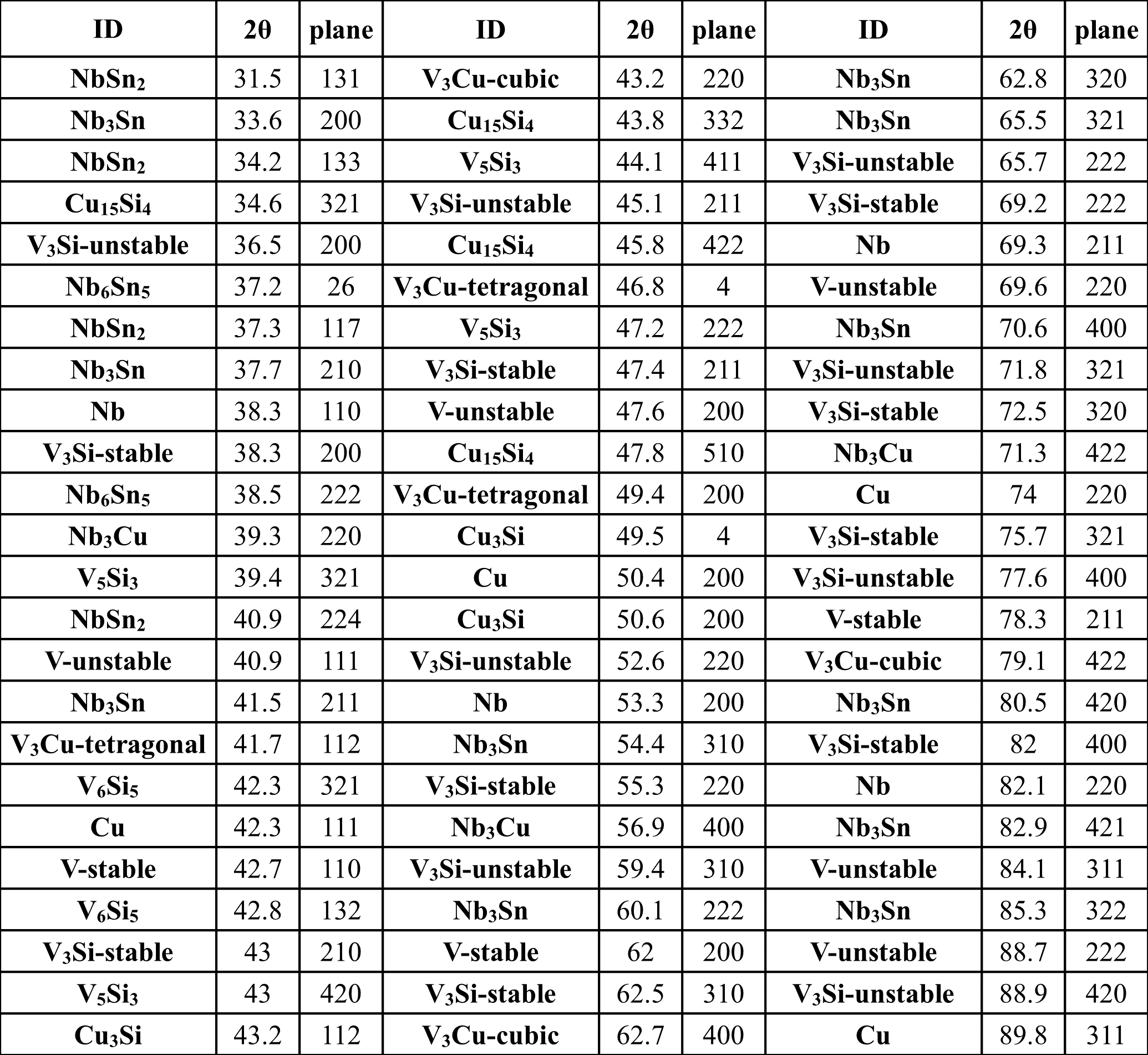}
\end{figure*}

\begin{figure*}[!h]
    \centering
    \renewcommand\figurename{Table}
    \caption{Strain for all detected peaks compared to known peak locations.}
    \includegraphics[width=\linewidth]{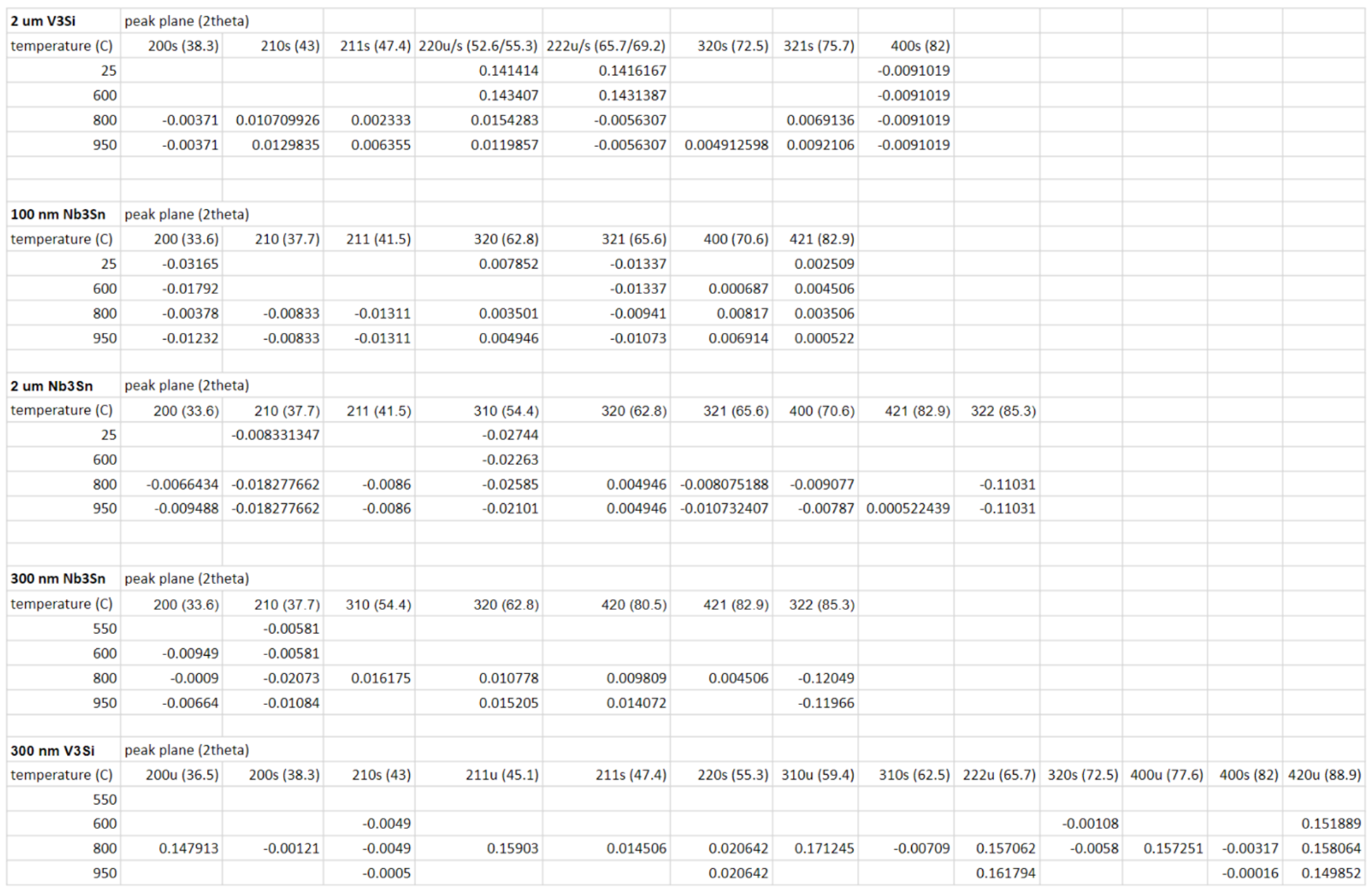}
\end{figure*}

\end{document}